\documentclass{article}

\input{tcilatex}

\begin{document}

\section{Contextual Observables and Quantum Information.}

\begin{center}
\textbf{M.Kupczynski\medskip }

\ Department \ of Mathematics and Statistics , Ottawa University\medskip
\end{center}

\begin{quotation}
\textbf{Abstract.} In this short paper we present the main features of a new
quantum programming language proposed recently by Peter Selinger which gives
a good idea about the difficulties of constructing a scalable quantum
computer. We show how some of these difficulties\ are related to the
contextuality of quantum observables and to the abstract and statistical
character of quantun theory (QT). We discuss also, in some detail, the
statistical interpretation (SI) of QT and the contextuality of observables
indicating the importance of these concepts for the whole domain of quantum
information.

\bigskip

\textbf{Keywords:}Entanglement , qubits, quantum information, quantum
computer , quantum measurement, quantum cryptography ,foundations of quantum
theory, contextual observablescontextual observables.
\end{quotation}

\textbf{PACS Numbers}: 03.65. Ta, \ \ \ 03.67.Lx \ , 03.67. Dd, \ \ 03.67.Hk

\subsection{\protect\bigskip Introduction}

\noindent In classical information theory a fundamental unit is a bit taking
the values 0 or 1 which can be easily implemented by presence or absence of
a particular physical signal. In quantum information theory a fundamental
unit is a qubit whose values are vectors in a Hilbert space H$_{1}$=C$^{2}$
which are in one-to-one correspondence with the states of some two level
physical system. Using two bits we may send 4 messages : 00, 01, 10, 11.
With two qubits the messages are the vectors in H$_{2}$=C$^{2}\otimes $C$%
^{2} $ \ with n qubits they are the vectors in H$_{n}$. This richeness of
the information together with long range correlations of the entangled
quantum states stimilated a vigorous research on the quantum information
theory and in particular on the quantum computer project.$\left[ 10,15,30%
\right] $

The quantum computation has been usually studied in terms of the Turing
machines , or in terms of circuits and gates. Several detailed quantum
programming languages have been recently developped$\left[ 7,23,24,31,38%
\right] $ . In particular an attractive functional programming language was
proposed by Selinger$\left[ 35\right] .$ With all these developments
understand better what kind of the quantum hardware is needed . This quantum
device(computer) has to manipulate the states of the qubits in a well
defined rapid and precise way. These states have to to be implemented as the
states of a particular quantum system. There are several proposed physical
implementations of qubits and quantum gates using: trapped ions$\left[ 12,22%
\right] ,$ cavity QED$\left[ 40\right] ,$ nuclear spins$\left[ 21\right] $,
Josephson junctions$\left[ 32\right] $ and quantum dots$\left[ 17,29,41%
\right] .$ The states of interest of the quantum register are all pure
states of n qubits. To make a quantum computer,with error correcting
ability, the quantum register should contain from 10$^{4}$ to 10$^{6}$
qubits . One should be able to to perform any unitary operation on the
qubits using an appropriate sequences of some universal set of quantum gates 
$\left[ 14,30,41\right] $ . One should be able to make a measurement in some
chosen common measurement basis on each chosen qubit without disturbing the
neighbouring qubits. The interactions of the system with the environment
should be controlled in order to avoid a decoherence of the created pure
states$\left[ 4,5,20\right] $. Finally the system should be scalable it
means that the difficulty of performing gates , measurements , etc., should
not grow too fast with the number of qubits in the registerin this short. In
spite of the magnificent technological progress in nanophysics$\left[ 17,22%
\right] $ and optics $\left[ 10,28,42\right] $we are still far from the
engineering stage of the quantum computer and the goal of 10$^{-5}$ errors
in the functioning of the gates is far from being achieved experimentally$%
\left[ 39,20\right] $ . It is relatively easy task, on a paper, to
maniputate entangled states of the qubits, to switch on and off the
interactions, to apply a sequence of the unitary operators or to perform a
particular measurement leading to the instantenuous collapse of the quantum
state. To do it meaningfully and in a controllable way in the experiment is
much more difficult because the quantum states are not the attributes of the
individual quantum systems.

The states of the classical objects are characterized by well defined
attributes which may be changed in the deterministic way quasi
instantaneously . The states of the quantum system are described by the
vectors $\Psi $ or by a density operator $\varrho $ which give the
statistical predictions for the repeated measurements\ of quantum
observables performed after the identical preparatory manipulations on the
same or on \ the ''identical'' quantum systems.$\left[ 8,9,16\right] .$

The problem is that our knowledge about the microworld is always indirect $%
\left[ 11,25\right] $ and that we may only estimate the values of the
physical observables from the empirical distributions of the experimental
results. No experimental result in the microworld may be predicted with the
certainty $\left[ 9,27\right] $and the quantum theory supplemented sometimes
by some stochastic ad hoc assumptions gives only the predictions on the
statistical regularities observed in the experiments. This applies as well
toall classic quantum experiments as to the experiments studying the
interference effects using low intensity sources of entangled photons $\left[
33,34,42\right] $ or beams of heavy C$_{60}$ molecules$\left[ 6\right] $.
This applies also to the experiments with the trapped ions and the quantum
dots.

It seems to us that any successful device in quantum information must take
into consideration the fact that the values of all quantum observables are
contextual and the manipulation of the quantum state has nothing to do with
the change of the attributes of some well defined and localised physical
microsystem.

The paper is organized as follows. In the section 1 we give a short
discussion of the attributive and contextual observables in physics. In the
section 2 we recall a formalism and the statistical interpretation of the
quantum theory.In the section 3 we talk about qubits and quantum gates. In
the section 4 we present a principle features of the Selinger's quantum
programming language.In section 5 we mention few implementation problems.

\subsection{\protect\bigskip Contextual observables}

Everyday observations taught us that the classical objects can be described
and classified according to their characteristic properties, called
attributes, such as their: form, size, dimension, colour etc.We found the
effective classifications of plants, animals, chemical substances, stars and
elementary particles. The static attributes are not sufficient to
differentiate between different behaviour of the physical objects in various
experiments this is why the contextual properties have been introduced.

An attributive property as a constant property of an object which can be
observed and measured at any time and which is not modified by the
measurement.

A contextual property as a property revealed only in a specific experiment
and characterizing the interaction of the object with the measuring
apparatus. Let us quote Accardi's example of a colour of a chameleon which
is green on the leaf and brown on the bark of the tree what ressembles the
behaviour of quantum systems $\left[ 1,2\right] $.

Another important \ example of contextual properties , inspired by some
random experiments we discussed in [26], is the following. Let us imagine
that we have several double sided coins \ C$_{1}$,..C$_{k}$ having two faces
: red (''R'') and green (''G'') . The coin my be flipped by different
flipping devices D$_{1}$...D$_{n}$. Each time if we flip a coin C$_{i}$
using the device D$_{j}$ we assume that we have an independent random
Bernoulli trial with a probability of a success: p(R)=p$_{ij}$. \ If p$_{ij}$%
. $\neq $p$_{kl}$ if i$\neq $k and j$\neq $l we clearly see that the
probability p$_{ij}$ is neither an attribute of the coin C$_{i}$nor the
attribute of the device D$_{j}$ . Each value p$_{ij}$ may be called a
contextual property of the coin C$_{i}$. In general the values p$_{ij}$ are
not known in advance and they can only be estimated within the error bars
from the the long runs of the corresponding random experiments.

A much more detailed discussion of the contextual observables may be found
in [25] where we considered \ the sources of some hypothetical particle
beams, detectors( counters), filters,transmitters and instruments and we
obtained the following general results.:

1) Properties of the beams depend on the properties of the devices and
vice-versa and are defined only in terms of the observed interactions
between them. For example a beam \ b is characterized by the statistical
distribution of outcomes obtained by passing by all the devices d$_{i}$. A
device d is defined by the statistical distributions of the results it
produces for all available beams b$_{i}$. All observables are contextual and
physical phenomena observed depend on the richness of the beams and of the
devices.

2) In different runs of the experiments we observe the beams b$_{k}$ each
characterized by its empirical probability distribution. Only if an ensemble
\ss\ of all these beams is a pure ensemble of pure beams we can associate
the estimated probability distributions of the results with the beams b$\in $%
\ss\ and with the individual particles members of these beams.

3) A pure ensemble \ss\ of pure beams b is characterized by such probability
distribution s(r) which remains approximately unchanged:

(i) for the new ensembles \ss $_{i}$ obtained from the ensemble \ss\ by the
application of the i-th intensity reduction procedure on each beam b$\in $\ss

(ii) for all rich sub-ensembles of \ss\ chosen in a random way

The quantum theory gives the probabilistic predictions thus it provides only
a contextual description of the physical systems.

There is a wholeness in any physical experiment$\left[ 8,9\right] $ $.$The
initial states are prepared and calibrated, they interact with the
experimental arrangement and the modified final states and/or the final
numerical results the measurements are found. The quantum theory (QT) does
not give any intuitive spatio-temporal picture of what is physically
happening ,the QT gives only the predictions about the final states and
about statistical distribution of the counts of the detectors.

\subsection{\protect\bigskip Quantum formalism.}

Let us recall now a standard description of the pure state of a quantum
system \ with infinite number of degrees of freedom We consider a Gelfand
triplet of spaces $\Omega \subset H\subset \Omega ^{\prime }$ where $\Omega $
is for example a Schwartz space of rapidly decreasing and infinitely
differentiable functions on R$^{n}$, H=L$^{2}$(R$^{n}$) and $\Omega ^{\prime
}$ is a space of tempered distributions on $\Omega .$ Let us consider two
observables A and B measured in the mutually exclusive experiments I and II
which are represented by self adjoint non commuting operators $\widehat{A}$
\ and $\ \widehat{B}.$ If $\ \ \left\{ \varphi _{\lambda }^{A}\right\}
_{\lambda \in \Lambda }$ and $\left\{ \varphi _{\gamma }^{B}\right\}
_{\gamma \in \Gamma }$ are complete sets of the generalised eigenfunctions (
tempered distributions) of $\widehat{A}$ \ and $\ \widehat{B}$ then we get
two different eigenvalue expansions of the unit state vector $\Psi \in H$ of
a studied physical system:

\bigskip $\Psi =\stackunder{\Lambda }{\int }\psi _{_{A}}(\lambda )\varphi
_{\lambda }^{A}d\lambda \ \ $\ \ \ \ \ \ \ \ \ \ \ \ \ \ \ \ \ \ \ \ \ \ \ \
\ \ \ (1)

\bigskip and

$\Psi =\stackunder{\Gamma }{\int }\psi _{_{B}}(\gamma )\psi _{\gamma
}^{B}d\gamma $\ \ \ \ \ \ \ \ \ \ \ \ \ \ \ \ \ \ \ \ \ \ \ \ \ \ \ \ \ (2)

\bigskip

where \ $\psi _{_{A}}(\lambda )$ = $\left\langle \varphi _{\lambda
}^{A},\Psi \right\rangle $ and \ $\psi _{_{B}}(\gamma )$ = $\left\langle
\varphi _{\gamma }^{B},\Psi \right\rangle $ are square integrable complex
value functions. The probabilities that the measured value of A in the
experiment I will fall into the interval [a,b] and that the measured value
of B in the experiment II will fall in the interval [c,d] are given by the
following well known formulae:

P(a$\leq $A$\leq $b$)=$ $\int\limits_{a}^{b}\lambda \overline{\psi
_{_{A}}(\lambda )}$ $\psi _{_{A}}(\lambda )d\lambda $ ,\ \ \ \ \ \ \ \ \ (3)

\bigskip P(c$\leq $B$\leq $d$)=$ $\int\limits_{c}^{d}\gamma \overline{\psi
_{_{B}}(\gamma )}$ $\psi _{_{B}}(\gamma )d\gamma $ .\ \ \ \ \ \ \ \ \ \ (4)

If a state of the system changes in time $\Psi (t)=U(t)\Psi ,$ where U(t) is
a unitary time evolution operator. The formulae (1) and (2) are
mathematically equivalent but they describe completely different
experiments. The predictions of QT for any experiment measuring the values
of the observable O are always obtained using a couple ( a state $\Psi $ and
an operator $\widehat{O}$ ) and clearly all the information about the
physical system is contextual and probabilistic. A value of a physical
observable is not an attribute of the system but it is the characteristic of
the pure ensemble of the experimental results created in the interaction of
the system with the experimental device.

The fact that using the same state vector $\Psi $ we may obtain the
predictions on the results of all different measurements performed on the
system led to the statements that the QT provides a complete description of
each individual physical system. From this statement there is a one step to
treat a state vector $\Psi $ as an attribute of each single individual
system what is incorrect and what leads to false paradoxes$\left[ 9,16,27%
\right] $.

The most states in physics are not pure but mixed. In the case of a m-level
system the most suitable is the formalism of the density operator $\widehat{%
\rho }$ acting in m dimensional hilbert space H$_{m}$ such that the
expectation value of of any observable O is given by

\bigskip $\left\langle O\right\rangle =$Tr( $\widehat{\rho }\widehat{O})$ \
.\ \ \ \ \ \ \ \ \ \ \ \ \ \ \ \ \ \ \ \ \ \ \ \ \ \ \ \ \ \ \ \ \ \ \ \ \ \
\ \ \ (5)\ \ \ \ \ \ \ \ 

The time evolution of $\widehat{\rho }(t)$ is given by:

$\widehat{\rho }(t)=U(t)\widehat{\rho }U(t)^{\ast }$\bigskip\ \ \ \ \ \ \ \
\ \ \ \ \ \ \ \ \ \ \ \ \ \ \ \ \ \ \ \ \ \ \ \ \ \ (6)

where * denotes the hermitian conjugation.

The density operator formalism is more general since by coupling a system to
the environment one may describe in terms of the reduced dynamics$\left[ 3,19%
\right] $ the passage from the pure to the mixed states allowing a
description of the decoherence or of the measurement process. One gets
simpler formalism by representing the operator $\widehat{\rho }$ by m\ $%
\asymp $ m positive hermitian matrix $\rho .$ The density matrices and their
transformations are building blocks of the quantum programming language$%
\left[ 35\right] $. In the follownig two sections we present the elements of
this language.

\subsection{\protect\bigskip Qubits and quantum gates}

A pure state of a single qubit is a formal linear combination of two known
states of some computational basis:q=$\alpha \mathbf{0}+\beta \mathbf{1}$ \
\ which may be represented by a unit column vector \textbf{u} in C$^{2}$ and
/or by 2$\times 2$ complex density matrix \textbf{u u* . }The n qubit states
are spanned by the tensor products of n single qubit states and all their
states pure and mixed may be represented by 2$^{n}\times $2$^{n}$ complex
density matrices M$.$.

A unitary operations S on n qubit states are called n-ary quantum gates and
their action on the state is defined by equation (6): SMS*.

If S is n-ary quantum gate, then the corresponding \textit{controlled }gate S%
$_{c}$ \ is the n+1- ary gate defined by \ the 2$^{n+1}\times $2$^{n+1}$
matrix::\bigskip

S$_{c}$= $\left[ \ 
\begin{tabular}{ll}
I & 0 \\ 
0 & S
\end{tabular}
\right] $ \ \ \ \ \ \ \ \ \ \ \ \ \ \ \ \ \ \ \ \ \ \ \ \ \ \ \ \ \ \ \ \ \
\ \ \ \ \ (7)

\bigskip

where I and 0 are 2$^{n}\times 2^{n}$identity matrix and zero matrix
respectively . In general a matrix with square brackets will be a matrix
composed of 4 blocks each containing square matrices of the same dimensions.

The following set of four unary and 5 binary gates can be chosen to be built
into the hardware$\left[ 35\right] $:

\bigskip

N=$\left[ \ 
\begin{tabular}{ll}
0 & 1 \\ 
1 & 0
\end{tabular}
\right] $ \ ,\ \ \ \ H=$\frac{1}{\sqrt{2}}\left[ \ 
\begin{tabular}{ll}
1 & 1 \\ 
1 & -1
\end{tabular}
\right] $ \ ,\ \ \ \ \ V=$\left[ \ 
\begin{tabular}{ll}
1 & 0 \\ 
0 & i
\end{tabular}
\right] $\ ,\ \ \ W= $\left[ \ 
\begin{tabular}{ll}
1 & 0 \\ 
0 & $\sqrt{i}$%
\end{tabular}
\right] $ ,

\bigskip

N$_{c}$, \ \ \ \ \ H$_{c}$, \ \ \ \ V$_{c}$, \ \ \ \ W$_{c}$, \ \ \ \ and \
\ \ X=$\left[ 
\begin{tabular}{llll}
1 & 0 & 0 & 0 \\ 
0 & 0 & 1 & 0 \\ 
0 & 1 & 0 & 0 \\ 
0 & 0 & 0 & 1
\end{tabular}
\right] .$ \ \ \ \ \ \ \ \ \ \ \ (8)

\bigskip

A set is not unique$\left[ 30,41\right] $ but it is sufficient to
approximate the action of any unitary gate according to the theorem $\left[
14\right] $which says that for any unitary matrix

S$\in $ C$^{2^{n}\times 2^{n}}$ and $\epsilon $ $\rangle $ $0$ there exist a
unitary matrix S' and a unit complex number $\lambda $ such that

$\Vert $S-$\lambda $ S'$\Vert $\TEXTsymbol{<}$\epsilon $ \ \ \ \ \ \ \ \ \ \
\ \ \ \ \ \ \ \ \ \ \ \ \ \ \ \ \ \ \ \ \ \ \ \ \ \ \ \ \ \ \ \ \ \ \ \ \ \
\ \ \ \ \ \ \ \ \ \ \ \ \ \ \ \ \ \ \ \ \ \ \ \ \ \ \ \ \ \ \ \ \ \ \ \ \ \
\ (9)\ \ \ \ 

\ \ \ \ \ \ \ \ \ \ \ \ \ \ \ \ \ \ \ \ \ \ \ \ \ \ \ \ 

and such S'=I$\otimes $A$\otimes $J where I,J are identity matrices of the
appropriate dimensions, and A is one of the gates H, V$_{c}$ or X.

\bigskip

In supplement to the unitary reversible operations, the operation ''
measure'' is introduced:

\bigskip

\ \ $\left[ 
\begin{tabular}{ll}
A & B \\ 
C & D
\end{tabular}
\right] \rightarrow \left[ 
\begin{tabular}{ll}
A & 0 \\ 
0 & 0
\end{tabular}
\right] or\left[ 
\begin{tabular}{ll}
0 & 0 \\ 
0 & D
\end{tabular}
\right] $ $\ \ \ \ \ \ \ \ \ \ \ \ \ \ \ \ \ \ \ \ \ \ \ \ \ \ \ \ \ \ \ \ \
\ $(10)

\bigskip

where the first matrix corresponding to a classical bit 0 is obtained with
the probability Tr A and the second corresponding to the classical bit 1 is
obtained with the probability Tr B . If the classical bits of information
are ignored the final result is a normal sum of the outcome matrices(10):

\bigskip\ $\left[ 
\begin{tabular}{ll}
A & B \\ 
C & D
\end{tabular}
\right] \rightarrow \left[ 
\begin{tabular}{ll}
A & 0 \\ 
0 & D
\end{tabular}
\right] $ \ \ \ \ \ \ \ \ \ \ \ \ \ \ \ \ \ \ \ \ \ \ \ \ \ \ \ \ \ \ \ \ \
\ \ \ \ \ \ \ \ \ \ \ \ \ \ \ \ \ \ \ \ \ (11)

\bigskip

\subsection{\protect\bigskip Quantum programming languages}

Several proposals for the quantum programming languages has been made. Let
us mention here the pseudo-cod formulation by Knill $\left[ 23,24\right] $,
Omer's rich procedural QCL$\left[ 31\right] $, Bettelli's quantum C++ $\left[
7\right] $and Sanders and Zuliani qGCL $\left[ 38\right] .$All these quantum
programming languages are so called imperative type languages because the
quantum data are manipulated in terms of arrays of quantum bit references
which requires the insertion of a number of run-time checks into the
compiled code which must be executed at run- time not at the compile time.

The quantum programming language proposed by Selinger$\left[ 35\right] $ is
a functional programming language with a static type system which guarantees
the absence of any run-time errors. The syntax and semantics of the language
allow high-level features such as loops, recursive procedures, and
structured data types. Each statement operates by transforming a specific
set of inputs to outputs. and the principle of non-cloning of quantum data
is enforced by the syntax.Both data flow and control flow are described
using the paradigm ``quantum data, classical control'' in consistence with
Shor's factoring algorithm, Grover's search algorithm and the Quantum
Fourier Transform $\left[ 13,14,30,36,37\right] $.

The models contains the classical and quantum flow charts . Quantum flow
charts are similar to classical flow charts, except that a new type \textbf{%
qbit} of quantum bits, and two new operations:unitary transformations and
measurements are added.. The following notation is used for these new
operations:\ q$\ast $=S \ \ for application of a unary unitary quantum gate
S to a quantum bit q, \ p,q$\ast $=S \ for a binary quantum gate S applied
to a pair p ,q\ of quantum bits,\TEXTsymbol{<}measure p\TEXTsymbol{>}: \ \
for the branching statement giving a couple of outcomes after the
measurement of p, ''$\circ $'' for the merge operation yielding the final
mixed state etc.

In an an example taken from $\left[ 35\right] $ a flow chart below
corresponds to the fragment of the program which: inputs two quantum bits p
and q, measures p, and then performs one of two possible unitary
transformations depending on the outcome of the measurement.The output is
the modified pair p,q.

\ \ \ \ \ \ \ \ \ \ \ \ \ \ \ \ \ \ \ \ \ \ \ \ \ \ \ \ \ \ \ \ \ \ \ \ \ \
\ \ input p,q: \textbf{qbit}

\ \ \ \ \ \ \ \ \ \ \ \ \ \ \ \ \ \ \ \ \ \ \ \ \ \ \ \ \ \ \ \ \ \ \ \ \ \
\ \ \ \ \ \ \ \ \ \ \ \ \ $\downarrow $

\ \ \ \ \ \ \ \ \ \ \ \ \ \ \ \ \ \ \ \ \ \ \ \ \ \ \ \ \ \ \ \ \ \ \ \ \ \
\ \ \ \ \ \ \ measure p

\ \ \ \ \ \ \ \ \ \ \ \ \ \ \ \ \ \ \ \ \ \ \ \ \ \ \ \ \ \ \ \ \ \ \ \ \ \
\ \ \ \ \ {\Large \ \ }{\small 0}$\swarrow ${\Large \ }$\ \searrow ${\small 1%
}

\ \ \ \ \ \ \ \ \ \ \ \ \ \ \ \ \ \ \ \ \ \ \ \ \ \ \ \ \ \ \ \ \ \ \ \ q$%
\ast $=N \ \ \ \ \ \ \ \ \ \ p$\ast $=N\ \ \ \ \ \ \ \ \ \ \ \ \ \ \ \ \ \ \
\ \ \ \ (11)

\ \ \ \ \ \ \ \ \ \ \ \ \ \ \ \ \ \ \ \ \ \ \ \ \ \ \ \ \ \ \ \ \ \ \ \ \ \
\ \ \ \ \ \ \ \ \ $\searrow $ \ \ $\swarrow $\ \ \ \ \ 

\ \ \ \ \ \ \ \ \ \ \ \ \ \ \ \ \ \ \ \ \ \ \ \ \ \ \ \ \ \ \ \ \ \ \ \ \ \
\ \ \ \ \ \ \ \ \ merge\ ($\circ )$

\ \ \ \ \ \ \ \ \ \ \ \ \ \ \ \ \ \ \ \ \ \ \ \ \ \ \ \ \ \ \ \ \ \ \ \ \ \
\ \ \ \ \ \ \ \ \ \ \ \ \ {\Large \ }$\downarrow $

\ \ \ \ \ \ \ \ \ \ \ \ \ \ \ \ \ \ \ \ \ \ \ \ \ \ \ \ \ \ \ \ \ \ \ \ \ \
\ \ \ output p,q: \textbf{qbit}\ \ 

\ \ \ \ \ \ \ \ 

If the input is a general mixed state \ described by a 4$\times 4$ density
matrix M written in the block notation the flow chart describes the
following sequence of operations performed on the density matrix M.

\bigskip

$\left[ 
\begin{tabular}{ll}
A & B \\ 
C & D
\end{tabular}
\right] $\ $\rightarrow $ $\left[ 
\begin{tabular}{ll}
A & 0 \\ 
0 & 0
\end{tabular}
\right] ,$ $\left[ 
\begin{tabular}{ll}
0 & 0 \\ 
0 & D
\end{tabular}
\right] \rightarrow $

$\bigskip $

$\left[ 
\begin{tabular}{ll}
NAN* & 0 \\ 
0 & 0
\end{tabular}
\right] ,$ $\left[ 
\begin{tabular}{ll}
D & 0 \\ 
0 & 0
\end{tabular}
\right] \rightarrow \left[ 
\begin{tabular}{ll}
A' & 0 \\ 
0 & 0
\end{tabular}
\right] $ $\ \ \ \ \ \ \ \ \ \ \ (12)$

\bigskip

where A' = NAN*+D . Let us note that by the usual mormalisation convention
the traces of all the density matrices are equal to the probabilities that
the corresponding edge is reached. If the input is a pure state, then the
states along each of the two branches of the measurement continue to be pure
and a final mixed state is obtained by the merge operation..

\subsection{Hardware requirements}

The quantum computing language could be implemented on various devices but a
real computation time gain could be only achieved using a quantum computer
par example a QRAM\ type machine$\left[ 23\right] $ which consists of a
classical computer which controls a special quantum hardware device such that%
$\left[ 35\right] $:

a) The quantum device provides a potentially large number of individually
addressable quantum bits.

b)Quantum bits can be manipulated via two fundamental operations: unitary
transformations and measurements.

c)The quantum device will implement a fixed, finite set of unitary
transformations which operate on one or two quantum bits at a time.

d)The classical controller communicates with the quantum device by sending a
sequence of instructions, specifying which fundamental operations are to be
performed.

e)The only output from the quantum device consists of the results of
measurements, which are sent back to the classical controller.

The operating system has to keep a list of quantum bits that are currently
in use by each process. When a process requests a new qubit, the operating
system finds a qubit that is not currently in use, marks that qubit as being
in use by the process, initializes the contents to\textbf{\ 0}, and returns
the addresse of the newly allocated qubit. The process can then manipulate
the qubit,for instance via operating system calls which take the qubit's
addressas as a parameter. The operating system ensures that processes cannot
access qubits that are not currently allocated to them -- this is very
similar to classical memory management. Finally, when a process is finished
using a certain qubit, it may deallocate it via another operating system
call; the operating system will then reset the qubit to \textbf{0} and mark
it as unused.In practice, there are many ways of making this scheme more
efficient,for instance by dividing the available qubits into regions, and
allocating and deallocating them in blocks, rather than individually.

Reseting or initializing a qubit to\textbf{\ 0 }is implemented by first
measuring the qubit, and then by performing a conditional ``not'' operation N%
$_{c}$ dependent on the outcome of the measurement.

In the introduction we reviewed various difficulties in constructing a
scalable quantum computing device. Let us add here some additional problems
related to the contextuality and SI of QT.

1)Concerning the measurements(10).Let us consider an entangled multiqubit
state$\Psi $ . The QT gives us statistical predictions on the results of
various long runs of coincidence type measurements which may be performed by
distant detectors on various qubits of a system.. For any particular qubit
we may obtain 0 or 1 with a given probability. Imagine now that in a given
moment of time a single measurement performed on the qubit q gives the
result 1 then according to SI we do not have any deterministic information
about the results of other coincidence measurements performed in the same or
future time on other qubits$\left[ 9,27\right] $. The collapsed state vector
gives us only \ the conditional probabilities for the the results of the
coincidence measurements performed on the remaining qubits corresponding to
subensemble of all n qubit events in which a measurement performed on the
qubit q gave 1.

2) Concerning the unitary gates (8,9). These transformations have to be
implemented by sequences of: qubit -external field and/or qubit-qubit
physical interactions. The ideal model requires an instataneous switching on
and off the various physical interactions without disturbing the other
qubits In practice one wants to use so called adiabatic switching which
seems to be very difficult task.

3) The produced pure states of quantum register have to be protected against
the decoherence due to the incontrollable influence of the environment.

\subsection{Conclusions{\protect\LARGE .}}

In our opinion any successful design of a quantum hardware for the purpose
of quantum information must be consistent with the contextual and
statistical character of quantum theory.

\subsection{References}

\begin{enumerate}
\item  Accardi L.,The quantum probabilistic approach to the foundationsof
quantum theory: urns and chamaleons,in in : ,M. Dalla Chiara, R. Giuntini,
F. Laudisa eds.,Philosophy of Science in Florence, Kluwer, Dordrecht, 1999
and Volterra Preprint N. 365 (1999)

\item  Accardi, L.and Regoli, M,, Locality and Bell's
Inequality,in:A.Khrennikov(ed.) QP-XIII, Foundations of Probability and
Physics, , World Scientific , Singapore,2002, 1--28

\item  Alicki, R.and Fannes, M, Quantum Dynamical Systems, Oxford Univ.
Press, Oxford, 2001.

\item  Alicki, R., Horodecki,M.,Horodecki,P.and Horodecki,R., Phys.Rev.A%
\textbf{\ 65}(2002),062101.

\item  Alicki, R.,A, Open.Sys. \& Information.Dyn.\textbf{11}(2004),53.

\item  Arndt, M.,Nairz,O., Vos-Andreae J., , Keller C., Van der Zouw, G. \&
Zeilinger, A., Nature\textbf{\ 401}(1999), 680682

\item  Bettelli, S.,Calarco,T. and Serani L.:Toward an architecture for
quantum programming, arXiv:cs.PL/0103009 v3, 2003

\item  Bohr,N. Essays 1958-1962 on Atomic Physics and Human Knowledge,
Wiley, New York,1963

\item  Ballentine, L. E.,Quantum Mechanics: A Modern Development ,World
Scientific , Singapore, 1998

\item  Bouwmeester,D.,Ekert,A.and Zeilinger,A.,The Physics of Quantum
Information, Springer-Verlag, Berlin, 2000

\item  Bush,P, Lahti,P.J and Mittelstaedt,P., The Quantum Theory of
Measurement, Lecture Notes in Physics, vol. m2 ,Springer, Berlin, 1991.

\item  Cirac, J.I, and Zoller, P., Nature \textbf{404} (2000),579.

\item  Cleve,R., Ekert,A.,.Macchiavello,C.and.Mosca,M., Proc. of the Royal
Soc.of London, A\textbf{454} (1998),339.

\item  Cleve.R., An introduction to quantum complexity theory,in
C.Macchiavello, G.Palma, and A.Zeilinger, eds, Collected Papers on Quantum
Computation and Quantum Information Theory, World Scientific,
Singapore,2000, 103.

\item  Deutsch D., Proc.of the Royal Soc. of London, A\textbf{400} (1985),97.

\item  Einstein, A., in:P.A Schilpp., (ed.), Albert Einstein:
Philosopher-Scientist, Harper \& Row, New York,1949

\item  Golovach, V.N. and Loss D., Semicond.Sci.Technol. \textbf{17} (2002)
355

\item  Grover,L.K Proc. of the 28th Annual ACM Symposium on the Theory of
Computing (STOC), (1996) pp.212-219 quant-ph/9605043

\item  Ingarden,R.,Kossakowski,A. and Ohya, M., Information Dynamics and
Open Systems,Kluver, Dordrecht, 1997

\item  Fonseca-Romero, K., Kohler, S, H\"{a}ngi, Chem.Phys.\textbf{296}%
(2004),307

\item  Kane,B.,Nature \textbf{393}(2000),133

\item  Kielpinski,D.,Monroe,C. and Wineland,D.J.,Nature \textbf{417} (2002),
709.

\item  Knill,E.H., Conventions for Quantum Pseudocode, unpublished, LANL
report LAUR-96-2724,1996

\item  Knill, E.H and Nielsen M.A, Theory of quantum computation,
quant-ph/0010057 (2001)

\item  Kupczynski,M. Int.J.Theor.Phys.\textbf{79}(1973), 319, reprinted in:
Physical Theory as Logico-Operational Structure,ed. C.A.Hooker,
Reidel,Dordrecht,1978,p.89

\item  Kupczynski, M., On the completeness of quantum mechanics ,
arXiv:quant-ph/028061 ( 2002)

\item  Kupczynski, M., Entanglement and Bell
inequalities,arXiv:quant-ph/0407199 (2004)

\item  Leonhardt,U.,Measuring the quantum states of light, Cambridge
Univ.Press, Cambridge, 1997

\item  Loss D. and DiVincenzo D.P.,Phys.Rev.A \textbf{57}(1998),120

\item  Nielsen,M.A. and Chuang.I.L.,Quantum Computation and Quantum
Information,Cambridge Univ.Press, Cambridge, 2000

\item  \"{O}mer B,. Quantum Programming in QCL, Master Thesis (2000),
http://tph.tuwien.ac.at/\symbol{126}oemer/qcl.html

\item  Pashkin, Y.A., Yamamoto,T., Astafiev,O.,Nakamura,Y.Averin, D.V., and
Tsai,,J.S,Nature \textbf{421}(2003),823.

\item  Rarity, J. , Phys. Rev. Lett. \textbf{65}(1990), 1348\ \ \ \ \ \ \ \
\ \ \ \ \ \ \ \ \ \ \ \ \ \ \ \ \ \ \ 

\item  Rarity, J.,Tapster,P., Phys. Rev.\textbf{45}, (1992,)2052

\item  Selinger P., Towards a quantum programming language, Mathematical
Structures in Computer Science (2004)\TEXTsymbol{\vert} in print.

\item  Shor,P.W., in S.Goldwasser.ed, Proceedings of the 35th IEEE FOCS,
IEEE, Los Alamitis, 1994,p.352 124--134, 1994..

\item  Shor,P.W., SIAM J. Computing \textbf{26} (1997),1484-1509.

\item  Sanders,J.W and Zuliani,P.,Quantum Programming, Math. of Program
Constr., Springer LNCS, 1837 (2000) pp.80-99

\item  Steane, A., Fortschr.Phys.\textbf{46}(1998),443.

\item  Turchette,Q.A et al., Phys. Rev.Lett., \textbf{75}(1995),4710

\item  DiVincenzo,D.P, Fortschr.Phys. \textbf{48} (2000),771

\item  Walther, P.,Pan, J-W.,Aspelmayer, M. Ursini,R.,Gasparoni S.and
Zeilinger A., Nature \textbf{429} (2004),158
\end{enumerate}

\end{document}